\documentclass{osa-article}

\journal{osac}

\articletype{Research Article}

\begin{document}

\title{High-Order Continuous-Variable Coherence of Phase-Dependent Squeezed State}

\author{Yanqiang Guo,\authormark{1,2,4} Haojie Zhang,\authormark{1,2} Xiaomin Guo,\authormark{1,2} Yuchi Zhang,\authormark{3} and Tiancai Zhang\authormark{3,5}}

\address{\authormark{1}Key Laboratory of Advanced Transducers and Intelligent Control System, Ministry of Education, College of Physics and Optoelectronics, Taiyuan University of Technology, Taiyuan 030024, China\\
\authormark{2}State Key Laboratory of Cryptology, Beijing 100878, China\\
\authormark{3}State Key Laboratory of Quantum Optics and Quantum Optics Devices, Institute of Opto-Electronics, Shanxi University, Taiyuan 030006, China}

\email{\authormark{4}guoyanqiang@tyut.edu.cn\\
\authormark{5}tczhang@sxu.edu.cn} 



\begin{abstract}
We study continuous variable coherence of phase-dependent squeezed state based on an extended Hanbury Brown-Twiss scheme. High-order coherence is continuously varied by adjusting squeezing parameter $r$, displacement $\alpha $, and squeezing phase $\theta $. We also analyze effects of background noise $\gamma $ and detection efficiency $\eta $ on the measurements. As the squeezing phase shifts from 0 to $\pi $, the photon statistics of the squeezed state continuously change from the anti-bunching ($g^{(n)}<1$) to super-bunching ($g^{(n)}>n!$) which shows a transition from particle nature to wave nature. The experiment feasibility is also examined. It provides a practical method to generate phase-dependent squeezed states with high-order continuous-variable coherence by tuning squeezing phase $\theta $. The controllable coherence source can be applied to sensitivity improvement in gravitational wave detection and quantum imaging.
\end{abstract}

\section{Introduction}

Quantum coherence characterizes inherent nature of light fields and reveals quantum statistics of photons. The research on quantum coherence originated from Hanbury Brown and Twiss (HBT) experiment \cite{HBT56} and subsequently, Glauber \cite{Glauber63} proposed a quantum theory that describes high-order coherence of light fields and makes it possible to distinguish between classical and non-classical light fields in quantum optics \cite{Kimble77,Davidovich96}. The non-classical feature \cite{Lee90,Filip18} of light fields is mainly reflected in anti-bunching \cite{Li19,Qu20}, squeezing \cite{Sun19,Kerdoncuff21} and entanglement \cite{Huo20}. In recent years, researchers have proved theoretically and experimentally existences of many non-classical optical states, such as Schr\"{o}dinger's cat state \cite{Omran19}, squeezed state \cite{Lee90,Sun19,Kerdoncuff21,Zhao20}, photon-added entangled coherent state \cite{Mojaveri14, Dehghani19, Dehghani20}, Fock state \cite{McKeever04,Guo12,Chu18,Yang19}. Among them, since squeezed state has an advantage of being lower than quantum shot-noise limit, its applications in gravitational waves detection \cite{McCuller20,Yu20}, quantum key distribution and optical communication \cite{Gehring15,Ruppert19,Pirandola20} have received extensive attention. Especially, squeezed state minimizes a combination of quantum radiation pressure noise and shot noise by tuning squeezing phase and squeezing level with a filter cavity to optimize sensitivity of Advanced LIGO detector and improve measurement precision \cite{McCuller20,Yu20}. As a typical non-classical light, continuous-variable squeezed light shows obvious photon bunching statistics (i.e. wave nature) \cite{Iskhakov12} and anti-bunching effects (i.e. particle nature) in special amplitude-displacement cases \cite{Koashi93,Grosse07}. Afterwards, many experimental methods for squeezed light generation with optical parameter oscillator (OPO) cavity \cite{Takanashi19}, four-wave mixing \cite{Slusher85,Marlow20}, and others \cite{Hosten16,Guo18} have been proposed, which has also promoted the research on the photon statistics of squeezed light. It is important to further the understanding of high-order coherence of squeezed light at single-photon detection level.

Second-order coherence $g^{(2)}$ is a standard metric for distinguishing different light fields. For example, when delay time $\tau =0$, $g^{(2)}>1$ indicates incoherence of light fields or indistinguishability of photons \cite{Sun17,Lan17,Guoy18,Zhou17,Su19}. $g^{(2)}=1$ indicates a coherent state and $g^{(2)}<1$ indicates a non-classical state with Sub-Poisson distribution which has anti-bunching statistics \cite{Wiersig09}. Accordingly, HBT scheme enables research on the second-order coherence of squeezed state with non-classical feature \cite{Iskhakov12,Koashi93}. In these works, the anti-bunching photon statistics of squeezed state stems from the interference of two-photon emission and coherent light, and the coherence can be partly changed by adjusting the intensity of the injected or interference light. The researches have been mainly devoted to elucidating the influences of squeezing parameter and displacement on the second-order coherence of squeezed state. However, the influence of squeezing phase on the coherence is not fully explored. Meanwhile, the second-order coherence $g^{(2)}$ only uncovers the variance of the photon number distribution. The higher-order coherences $g^{(3)}$ and $g^{(4)}$, which respectively reflect the skewness and kurtosis of the distribution, offer more information on multi-photon emission and can be used to characterize the nonclassical feature of light field \cite{Guo11,Guo20}. Extended HBT schemes combined with more single-photon detectors are used to access high-order coherences \cite{Guo20,Hamsen17}. The measurements of high-order coherences for different states have been investigated in the past decades \cite{Guo20,Hamsen17,Avenhaus10,Rundquist14}, and have been applied to ghost imaging \cite{Chen10,Zhou12}, quantification of timescales in phonon laser \cite{Xiao20}, characterization of single-photon detectors \cite{Wayne17}, and so on. The effects of squeezing phase, background noise, and detection efficiency on high-order coherences are also important for the quantum statistics of squeezed state. Controllable high-order coherences of squeezed state with comprehensive analysis of various influential factors, which continuously vary from bunching to anti-bunching, remain elusive and to be explored.

In this work, we exploit an extended HBT scheme with four single-photon counting modules (SPCMs) to investigate high-order coherence of phase-dependent squeezed state. High-order coherence of the squeezed state can be continuously varied from anti-bunching effect to super-bunching effect by adjusting the squeezing phase from 0 to $\pi $. The effects of squeezing degree, displacement, background noise and overall efficiency on the coherences are studied. The optimal anti-bunching and super-bunching effects are obtained with the same feasible parameters, except the squeezing phase. The continuous-variable coherence of the phase-dependent squeezed state is prepared for quantum information applications.

\section{Theoretical model}

\begin{figure}[htbp]
\centering
\includegraphics[width=\linewidth]{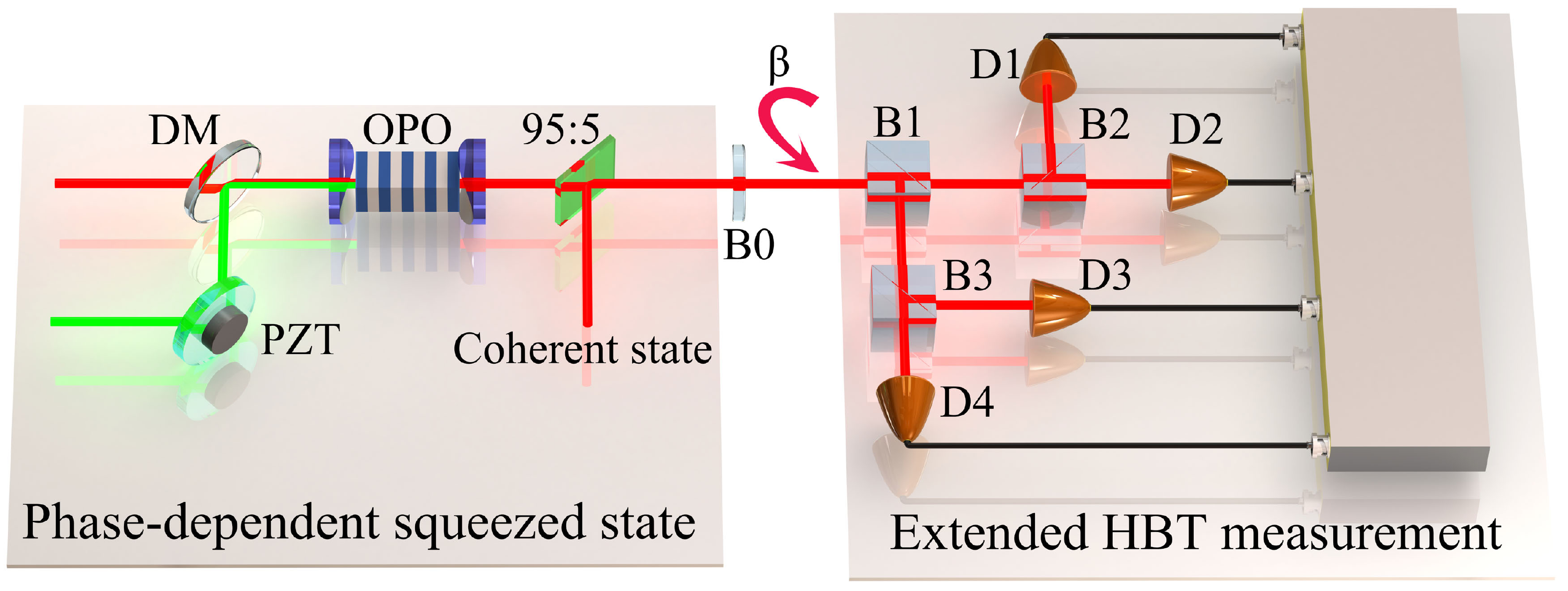}
\caption{Schematic diagram for examining high-order coherences of phase-dependent squeezed state based on extended HBT measurement using four SPCMs. $\eta $ is overall efficiency and $\beta $ is background noise. $B1$, $B2$, and $B3$ are 50/50 lossless beam splitters and $D1$, $D2$, $D3$, and $D4$ are single-photon counters. DM: dichroic mirror, PZT: piezoelectric ceramic transducer, OPO: optical parameter oscillator.}
\label{fig1}
\end{figure}

The schematic diagram is shown in Fig. \ref{fig1}. The input squeezed light is a non-classical photon source with oscillating photon number distributions \cite{Scully97}.
\begin{eqnarray}
P_{scs}(n) &=&\frac{\tanh ^{n}r}{2^{n}n!\cosh r}\exp \left[ -\left\vert
\alpha \right\vert ^{2}+\frac{1}{2}(e^{-i\theta }\alpha ^{2}+e^{i\theta
}\alpha ^{\ast 2})\tanh r\right]   \nonumber \\
&&\times \left\vert H_{n}(\frac{\alpha e^{-i\theta /2}}{\sqrt{2\cosh r\sinh r%
}})\right\vert ^{2},
\end{eqnarray}
where $\theta $ is the squeezing phase, $r$ is the squeezing parameter, and $\alpha $ ($\alpha \in R$) is the displacement. $H_{n}(x)$ is the Hermite polynomials.

We can prepare a phase-dependent squeezed state by the OPO cavity and measure the state by an extended HBT device. Then a beam splitter $B0$ with a transmittance $\eta $ is placed in front of the measurement unit to simulate the loss and detection efficiency of the entire system. Meanwhile, we also consider the effect of noise $\left\vert \beta \right\rangle $ on the system. The noise $\left\vert \beta \right\rangle $ is mainly caused by dark counts and background noise, and the photon number distribution follows the Poisson distribution $P_{noise}(n)=\gamma ^{n}\exp (-\gamma )/n!$, where $\gamma =\left\vert \beta \right\vert ^{2}$ represents the average photon number of noise. The noise-affected phase-dependent squeezed state then passes through three beam splitters $B1$, $B2$, $B3$, and finally the HBT detection system consisting of four SPCMs $D1$, $D2$, $D3$, and $D4$ measures its probability of joint photon number distribution, as shown in Fig. \ref{fig1}.

After the phase-dependent squeezed state with photon number distribution $P_{scs}(n)$ in Eq. (1) passes through the beam splitter $B0$ with the overall detection efficiency $\eta $, its photon number distribution evolves through the Bernoulli transformation as
\begin{equation}
\Lambda _{m}^{n}=\sum_{n=m}^{\infty }P_{scs}(n)\binom{n}{m}\eta ^{m}(1-\eta
)^{(n-m)},
\end{equation}
where $\binom{n}{m}=\frac{n!}{m!(n-m)!}$. The light field is then mixed with the background noise $\left\vert \beta \right\rangle $ at the beam splitter $B1$, which can be considered as a convolution process. The photon number distribution of the weak background noise $P_{noise}(n)$ obeys the Poissonian distribution. Therefore, the mixed photon number distribution arriving at $B1$ can be written as
\begin{equation}
\Lambda _{L}^{m}=\sum_{m=0}^{L}\gamma ^{(L-m)}e^{-\gamma }\Lambda
_{m}^{n}/\prod\limits_{i=1}^{L-m}i,
\end{equation}
where $L$ is the total photon numbers reaching the beam splitter $B1$. After $B1$, $N$ photons are transmitted and $L-N$ photons are reflected. Subsequently, the $N$ transmitted photons reach $B2$, and $K$ photons are reflected into detector $D1$, while $N-K$ photons are transmitted to detector $D2$. Then $L-N$ photons in the reflection path of $B1$ reach $B3$. $M$ and $L-N-M$ photons reach detectors $D3$ and $D4$ after reflection and transmission of $B3$, respectively.

The extended HBT device consisting of four SPCMs measures the joint photon probability $\Gamma _{click\otimes i}$, where $i$ indicates the number of clicking detectors. Since the four detectors have the same performance and the three beam-splitters $B1$, $B2$, and $B3$ in the double HBT system are all 50:50 beam splitters, the final photon probabilities are limited to five cases, i.e.
\allowdisplaybreaks
\begin{eqnarray}
\Gamma _{click\otimes 0} &=&\Lambda _{L=0}^{m},  \nonumber \\
\Gamma _{click\otimes 1} &=&4\sum_{L=1}^{\infty }\Lambda _{L}^{m}\left(
\frac{1}{2}\right) ^{2L},  \nonumber \\
\Gamma _{click\otimes 2} &=&6\sum_{L=2}^{\infty }\Lambda
_{L}^{m}\sum_{k_{1}=1}^{L-1}L!\left( \frac{1}{2}\right)
^{2L}/\prod\limits_{i=1}^{2}k_{i}!, \\
\Gamma _{click\otimes 3} &=&4\sum_{L=3}^{\infty }\Lambda
_{L}^{m}\sum_{k_{1}=2}^{L-1}\sum_{k_{3}=1}^{k_{1}-1}L!\left( \frac{1}{2}
\right) ^{2L}/\prod\limits_{i=2}^{4}k_{i}!,  \nonumber \\
\Gamma _{click\otimes 4} &=&\sum_{L=4}^{\infty }\Lambda
_{L}^{m}\sum_{k_{1}=2}^{L-2}\sum_{k_{3}=1}^{k_{1}-1}
\sum_{k_{5}=1}^{k_{2}-1}L!\left( \frac{1}{2}\right)
^{2L}/\prod\limits_{i=3}^{6}k_{i}!,  \nonumber
\end{eqnarray}
where $k_{1}=N$, $k_{2}=L-N$, $k_{3}=K$, $k_{4}=N-K$, $k_{5}=M$, and $k_{6}=L-N-M$. According to the definition of correlation function \cite{Glauber63}, the m-order coherence through photon counting measurement can be expressed as follows:
\begin{equation}
g^{(m)}=\frac{\left\langle n_{1}n_{2}\cdots n_{m}\right\rangle }{%
\left\langle n_{1}\right\rangle \left\langle n_{2}\right\rangle \cdots
\left\langle n_{m}\right\rangle }.
\end{equation}
where $n_{m}$ represents the photon number and $\left\langle \cdot \right\rangle $ denotes the ensemble average. We use the double HBT system to obtain the probability distribution $\Gamma _{click\otimes i}$ of the phase-dependent squeezed state and the high-order coherence is derived by substituting Eq. (4) into Eq. (5). Then we can have the second, third, and fourth-order coherences determined by the detected photon probabilities.
\begin{eqnarray}
g^{(2)} &=&\frac{8\Gamma _{click\otimes 2}+24\Gamma _{click\otimes
3}+48\Gamma _{click\otimes 4}}{3\left\langle n\right\rangle ^{2}}, \nonumber \\
g^{(3)} &=&\frac{16\Gamma _{click\otimes 3}}{\left\langle n\right\rangle ^{3}
}, \\
g^{(4)} &=&\frac{256\Gamma _{click\otimes 4}}{\left\langle n\right\rangle
^{4}}, \nonumber
\end{eqnarray}
where $\left\langle n\right\rangle=\sum_{i=0}^{\infty }i\Gamma _{click\otimes i}$ is the total average number of photons detected by the system. The detailed derivation of the high-order coherences based on the extended HBT scheme is given in Appendix.

\section{High-order coherence of phase-dependent squeezed state}

For an input phase-dependent squeezed state, it has the following expression:
\begin{equation}
\left\vert \xi ,\alpha \right\rangle =S\left( \xi \right) D\left( \alpha
\right) \left\vert 0\right\rangle ,
\end{equation}
where $S\left( \xi \right) $ and $D\left( \alpha \right) $ are unitary squeezing operator and displacement operator, respectively. $\xi =re^{i\theta }$ is the squeeze factor with the squeezing phase $\theta $ and $\alpha $ is the amplitude displacement. In the absence of delay time, the normalized high-order photon coherence can be obtained by using Eq. (5)
\begin{eqnarray}
g^{(2)} &=&1-\frac{1}{A^{2}}\left[ B-\left( 2\left\vert \Omega \right\vert
^{2}+\cosh 2r\right) \sinh ^{2}r\right], \nonumber \\
g^{(3)} &=&1-\frac{1}{A^{3}}\left[
\begin{array}{c}
3B\left( \left\vert \Omega \right\vert ^{2}+3\sinh ^{2}r\right) -\left(
2+7\cosh 2r\right) \sinh ^{4}r \\
-3\left\vert \Omega \right\vert ^{2}\sinh ^{2}r\left( 2\left\vert \Omega
\right\vert ^{2}+4\cosh 2r-1\right)
\end{array}
\right] , \\
g^{(4)} &=&1+\frac{1}{A^{4}}\left[
\begin{array}{c}
3B^{2}+\left( 3-7\cosh 2r+13\cosh 4r\right) \sinh ^{4}r+12\left\vert \Omega
\right\vert ^{6}\sinh ^{6}r \\
-6B\left( \left\vert \Omega \right\vert ^{4}-8\left\vert \Omega \right\vert
^{2}\sinh ^{2}r-3C\sinh ^{2}r\right)  \\
+6\left\vert \Omega \right\vert ^{2}\sinh ^{2}r\left( 2C+3\cosh 2r\right)
+4D\left\vert \Omega \right\vert ^{2}\sinh ^{4}r%
\end{array}%
\right], \nonumber
\end{eqnarray}
where $\Omega =\alpha \left( \cosh r-e^{i\theta }\sinh r\right) $, $A=\left\vert \Omega \right\vert ^{2}+\sinh ^{2}r$, $B=\left( \Omega ^{\ast 2}e^{i\theta }+\Omega ^{2}e^{-i\theta }\right) \cosh r\sinh r$, $C=\cosh 2r+3\sinh ^{2}r$, and $D=13\cosh ^{2}r+23\cosh 2r$. The equation (8) indicates that the ideal high-order coherences of the squeezed state are related to the three parameters, i.e. squeezing parameter $r$, amplitude displacement $\alpha $, and squeezing phase $\theta $. The $A$ and $B$ are both real numbers. Eq. (6) gives the high-order coherences with taking into account the effects of background noise and detection efficiency. If the input state is $\left\vert \alpha ,\xi \right\rangle =D\left( \alpha \right) S\left( \xi \right) \left\vert 0\right\rangle $ in which the displacing and squeezing operations are opposite to Eq. (7), the similar continuous-variable coherences are also obtained and it should be noted that the flexible operations provide powerful support for the experiment feasibility. The second-order coherence $g^{(2)}$ reflects the mean value of photon number distribution, and the third-order coherence $g^{(3)}$ and fourth-order coherence $g^{(4)}$ reflect the statistical skewness and kurtosis respectively. The analysis of the high-order coherences allows us to have a deeper understanding of quantum statistics of the squeezed state.

\section{Results}

\subsection{High-order coherence of phase-dependent squeezed state for $\theta =0$}

The high-order coherences of phase-dependent squeezed state is closely related to the three parameters: $r$, $\alpha $, and $\theta $. We investigate and obtain the high-order coherences of the phase-dependent squeezed state as functions of squeezing parameter $r$, displacement $\alpha $, and squeezing phase $\theta $. The experiment feasibility of this scheme is also verified.

\begin{figure}[htbp]
\centering
\includegraphics[width=0.95\linewidth]{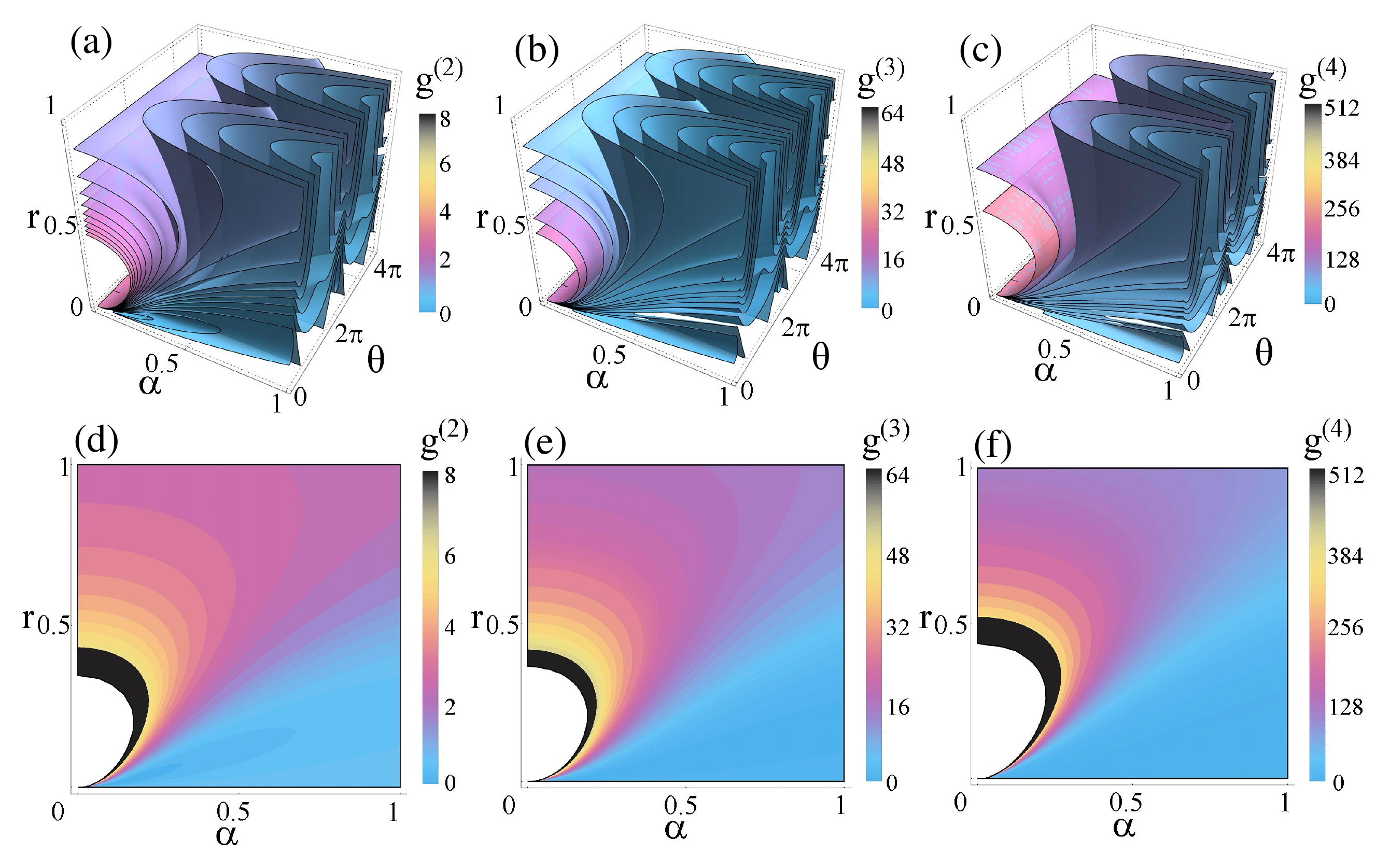}
\caption{(a)-(c) $g^{(2)}$, $g^{(3)}$, and $g^{(4)}$ versus squeezing parameter $r$, displacement $\alpha $, and squeezing phase $\theta $; (d)-(f) maps of $g^{(2)}$, $g^{(3)}$, and $g^{(4)}$ of the phase-dependent squeezed state versus $r$ and $\alpha $ when $\theta =0$.}
\label{fig2}
\end{figure}

Figure \ref{fig2} shows the results of high-order coherences versus the three parameters [Figs. \ref{fig2}(a)-\ref{fig2}(c)] according to Eq. (8), and the maps of high-order coherences versus $r$ and $\alpha $ when $\theta =0$ [Figs. \ref{fig2}(d)-\ref{fig2}(f)]. Light blue regions exhibit photon anti-bunching effect ($g^{(n)}<1$), and regions exhibiting photon super-bunching ($g^{(n)}>n!$) are marked progressively darker. In the ideal case, the high-order coherences $g^{(n)}$ give a continuous variation with the squeezing phase $\theta $, squeezing parameter $r$, and amplitude displacement $\alpha $, as shown in Figs. \ref{fig2}(a)-\ref{fig2}(c). When the amplitude displacement $\alpha $ is held constant, the $g^{(n)}$ exhibits a $2\pi $-periodic variation with the squeezing phase $\theta $ and can continuously shift from anti-bunching regions to super-bunching regions as the squeezing parameter $r$ increases. When the $r$ is kept constant, the $g^{(n)}$ also exhibits a $2\pi $-periodic variation with the $\theta $ and can continuously vary from super-bunching regions to anti-bunching regions as the $\alpha $ increases. Moreover, the photon statistics of the phase-dependent squeezed state can be changed continuously between anti-bunching effect ($g^{(n)}<1$) and super-bunching effect ($g^{(n)}>n!$) as the squeezing phase $\theta $ varies. When the phase $\theta $ is kept constant, the $g^{(n)}$ can continuously shift between super-bunching regions and anti-bunching regions as the displacement $\alpha $ and the squeezing $r$ vary. In Figs. \ref{fig2}(d)-\ref{fig2}(f), when the squeezing phase $\theta =0$ and the squeezing $r$ is weak, the $g^{(n)}$ can continuously vary from super-bunching effect to anti-bunching effect as the amplitude displacement $\alpha $ increases. For small $\alpha $ at $\theta =0$, the $g^{(n)}$ can continuously vary from anti-bunching regions to super-bunching regions as the r increases. The controllable coherence squeezed state is beneficial to boosting an implementation of high-speed, remote and scalable quantum communication, especially the communication of combining discrete-variable and continuous-variable approaches \cite{Pirandola20}. Moreover, the phase-dependent squeezed state with controllable coherence contributes to improving measurement precision of Advanced LIGO detection \cite{Yu20}. It is important and potentially useful for hybrid discrete- and continuous-variable quantum key distribution and high-precision quantum sensing.

\begin{figure}[htbp]
\centering
\includegraphics[width=0.9\linewidth]{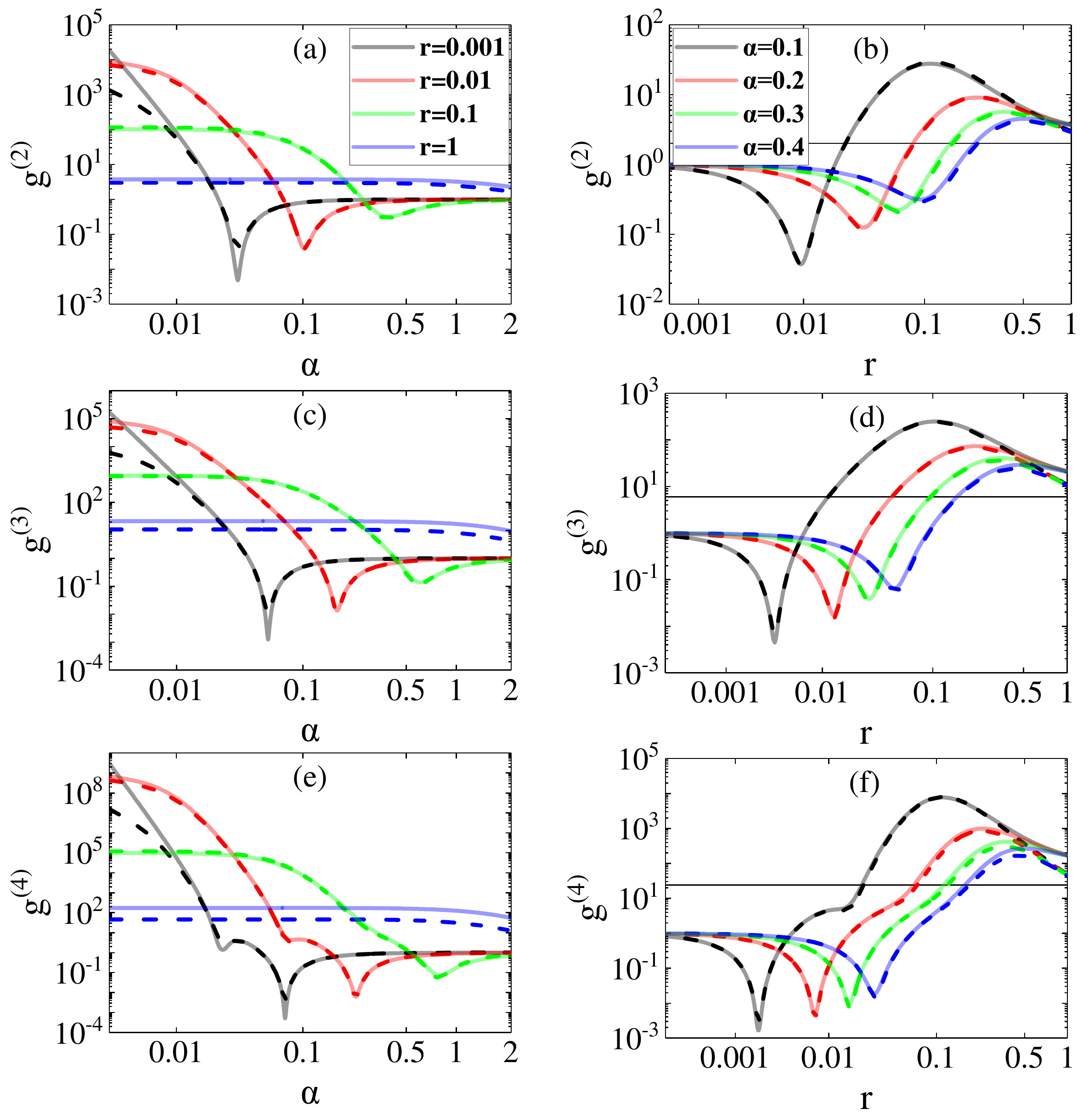}
\caption{High-order coherence $g^{(n)}$ ($n=2,3,4$) versus displacement $\alpha $ and squeezing parameter $r$ for squeezing phase $\theta =0$. The solid curves indicate the ideal case of background noise $\gamma =0$ and detection efficiency $\eta =1$. The dashed curves indicate the feasible case of $\gamma =10^{-5}$, $\eta =0.5$.}
\label{fig3}
\end{figure}

When $\alpha =0$, the phase-dependent squeezed state turns into a squeezed vacuum state. In our previous work, the extended HBT scheme has been used to specifically analyze the high-order coherences of the squeezed vacuum state versus the squeezing parameter and detection efficiency, and further details refer to \cite{Guo20}. When the displacement $\alpha \neq 0$, the high-order coherence $g^{(n)}$ ($n=2,3,4$) changes periodically with the phase $\theta $ on a cycle of $2\pi $. The phase-dependent squeezed state with squeezing phase $\theta =0$ can be prepared experimentally. Based on the double-HBT scheme, the theoretical results of Figs. \ref{fig3}(a), \ref{fig3}(c), \ref{fig3}(e) show that the high-order coherences follow a nonmonotonic dependence on the amplitude displacement $\alpha $, revealing a dip for a critical low $\alpha $. The dip indicates that two-photon destructive interference exists between the squeezing amplitude and the displacement amplitude. For small squeezing parameters $r$, the high-order coherences behave from super-bunching effect to anti-bunching effect, and then approach to 1, as the $\alpha $ increases. For the squeezing $r=0.001$, the minimum values of $g^{(2)}=0.0034$, $g^{(3)}=0.0009$, and $g^{(4)}=0.0003$ can be obtained when the $\alpha $ are 0.032, 0.055, and 0.074 respectively. The minimum anti-bunching values of $g^{(n)}$ increase as the squeezing $r$ increases. In Figs. \ref{fig3}(b), \ref{fig3}(d) and \ref{fig3}(f), the high-order coherences also show a nonmonotonic variation as the squeezing parameter $r$ increases, and the $g^{(n)}$ falls from 1 to the minimum anti-bunching values and continuously increases to super-bunching values. Meanwhile, as can be seen in Fig. \ref{fig3}, the theoretical results of $g^{(n)}$ become consistent with the ideal ones as the displacement $\alpha $ and squeezing parameter $r$ increase. The solid curves in Fig. \ref{fig3} correspond to the results obtained from Eq. (8) in an ideal case (i.e., $\eta =1$ and $\gamma =0$), and the dashed curves indicate the results obtained from Eq. (6) in a feasible experiment case ($\eta =0.5$ and $\gamma =10^{-5}$).

\begin{figure}[htbp]
\centering
\includegraphics[width=\linewidth]{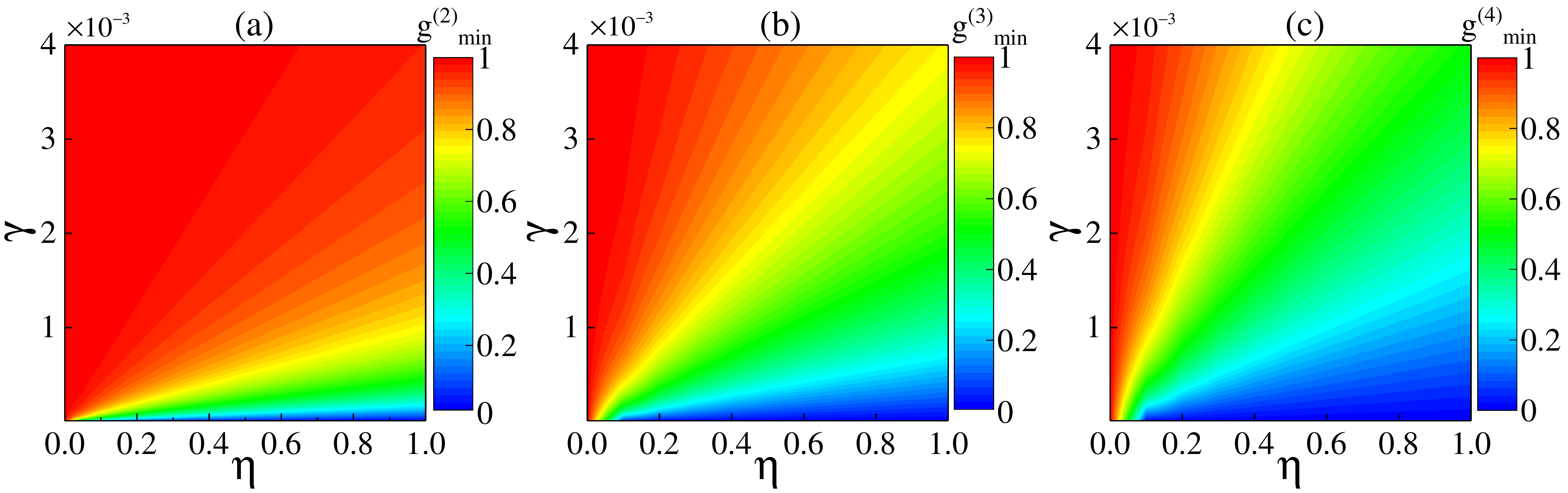}
\caption{Maps of (a) $g_{\min }^{(2)}$, (b) $g_{\min }^{(3)}$, and (c) $g_{\min }^{(4)}$ as functions of background noise $\gamma $ and detection efficiency $\eta $ when squeezing parameter $r=0.001$ and squeezing phase $\theta =0$.}
\label{fig4}
\end{figure}

Furthermore, it should be noted that due to the effects of background noise and detection efficiency, the $g^{(n)}$ deviation between the ideal values and the experimentally feasible values is relatively large for low displacement $\alpha $ and weak squeezing $r$. In Fig. \ref{fig4}, we analyze the effects of background noise $\gamma $ and detection efficiency $\eta $ on the minimum high-order coherences $g_{\min }^{(n)}$, which shows the strong anti-bunching effect. As the background noise $\gamma $ decreases and the detection efficiency $\eta $ increases, the $g_{\min }^{(n)}$ approaches to 0 when the squeezing parameter $r=0.001$ and squeezing phase $\theta =0$. Meanwhile, the anti-bunching effects of higher-order coherences $g^{(n)}$ ($n>2$) are more robust against the background noise $\gamma $ and the detection efficiency $\eta $ than that of $g^{(2)}$. The background noise $\gamma $ for $g^{(4)}<0.5$ is one order of magnitude larger than the background noise for $g^{(2)}<0.5$. It is easier to observe strong anti-bunching effect by measuring the higher-order coherence.

\subsection{High-order coherence of phase-dependent squeezed state for $\theta =\pi $}

\begin{figure}[htbp]
\centering
\includegraphics[width=\linewidth]{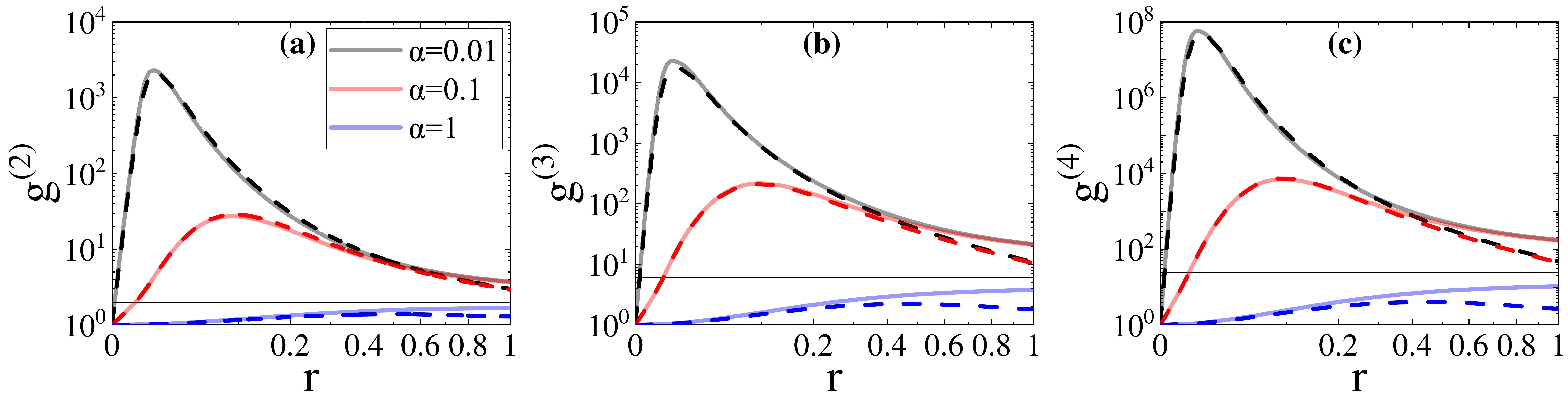}
\caption{High-order coherence $g^{(n)}$ as a function of squeezing parameter $r$ for squeezing phase $\theta =\pi $ with three $\alpha $: 0.01, 0.1, and 1. The solid curves indicate $g^{(n)}$ in an ideal case of $\gamma =0$, $\eta =1$ and the dashed curves indicate the results when $\gamma =10^{-5}$ and $\eta =0.5$.}
\label{fig5}
\end{figure}

To investigate the super-bunching effect of phase-dependent squeezed state, we also analyze the high-order coherence versus squeezing parameter $r$ for squeezing phase $\theta =\pi $, as shown in Fig. \ref{fig5}. The solid curves indicate the high-order coherence $g^{(n)}$ in the ideal case of background noise $\gamma =0$ and detection efficiency $\eta =1$, and the dashed curves indicate the $g^{(n)}$ when $\gamma =10^{-5}$ and $\eta =0.5$. The high-order coherence behaves super-bunching effect ($g^{(n)}>n!$) for small displacement $\alpha $, and the $g^{(n)}$ first increases to the maximum and then decreases as the squeezing parameter $r$ increases. For $\alpha =0.01$ and $\theta =\pi $, the maximum high-order coherences are $g^{(2)}=2.5\times 10^{3}$, $g^{(3)}=2.2\times 10^{4}$ and $g^{(4)}=5.6\times 10^{7}$ with $r=0.01$. The straight lines in Fig. \ref{fig5} represent the values of $n!$ ($n=2,3,4$). It should be noted that the high-order coherence $g^{(n)}$ can change from anti-bunching effect to super-bunching effect when the squeezing phase $\theta $ turns from 0 to $\pi $.

\subsection{$g^{(n)}$ of phase-dependent squeezed state for continuous $\theta $}

\begin{figure}[htbp]
\centering
\includegraphics[width=\linewidth]{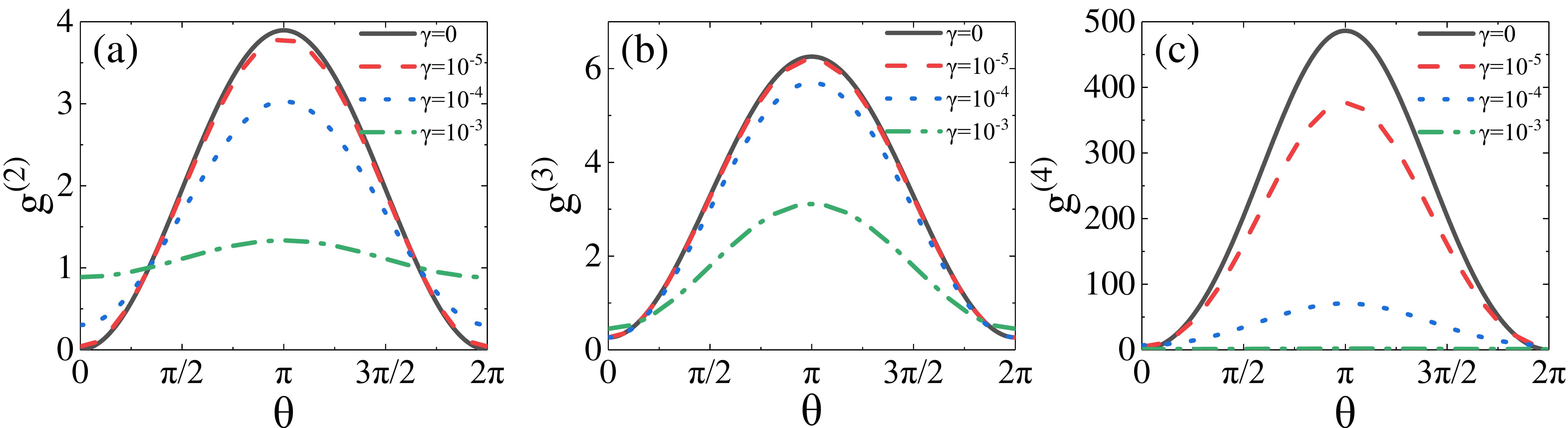}
\caption{$g^{(n)}$ as a function of squeezing phase $\theta $ for detection efficiency $\eta =0.5$ and various background noises $\gamma $.}
\label{fig6}
\end{figure}

The high-order coherences of phase-dependent squeezed state continuously vary from strong anti-bunching effect to super-bunching effect as the squeezing phase $\theta $ changes. The results are shown in Fig. \ref{fig6}. The $g^{(n)}$ versus $\theta $ varies with a period of $2\pi $, and the photon number distribution of phase-dependent squeezed state gradually tends to Poisson distribution as the background noise $\gamma $ increases. For the feasible detection efficiency $\eta =0.5$ and background noise $\gamma =10^{-5}$, the $g^{(2)}$ with $\alpha =0.032$ and $r=0.001$, $g^{(3)}$ with $\alpha =0.063$ and $r=0.002$, and $g^{(4)}$ with $\alpha =0.017$ and $r=5\times 10^{-4}$ allow us to observe the transition from anti-bunching effect ($g^{(n)}<1$) to super-bunching effect ($g^{(n)}>n!$) by continuously adjusting squeezing phase $\theta $.
It also indicates that the weak squeezing can induce the continuous-variable high-order coherences of phase-dependent squeezed state.

\section{Discussion}

With the development of single-photon detection technology, the typical SPCM detection efficiency with a range of $10\%$--$70\%$ can be achieved higher than $50\%$ at 852 nm. The count rate of background noise is from 25 counts/s to 1500 counts/s and the resolution time is higher than 350 ps. Accordingly, the background noise $\gamma $ can be as low as the order of $10^{-9}$ in the feasible experimental case. When $\theta =0$ with $\gamma =10^{-5}$ and $\eta =0.5$, the strong anti-bunching effects of $g^{(2)}=0.042$, $g^{(3)}=0.010$, $g^{(4)}=0.003$ are obtained at weak squeezing $r=0.001$ (i.e. 0.009 dB) and small displacements $\alpha =0.032$, $\alpha =0.055$, $\alpha =0.074$. The magnitude of squeezing in dB units is given by $-10\log _{10}e^{-2r}$. When $\theta =\pi $ with the same feasible parameters, the super-bunching effects of $g^{(2)}=3.786$, $g^{(3)}=6.190$, $g^{(4)}=375.9$ are obtained for $r=0.009$ dB and $\alpha =0.032$, $r=0.017$ dB and $\alpha =0.063$, $r=0.004$ dB and $\alpha =0.016$, respectively. The phase-dependent squeezed state can be prepared and its photon statistics shows a continuous variation from anti-bunching effect ($g^{(n)}<1$) to super-bunching effect ($g^{(n)}>n!$) as the squeezing phase $\theta $ is tuned. Furthermore, as the detection efficiency increases and the background noise decreases, the high-order coherences corresponding to the minimum anti-bunching and maximum super-bunching effects will be improved.

\section{Conclusion}

The high-order coherence of phase-dependent squeezed state based on the extended HBT scheme is investigated. The coherence versus the squeezing parameter $r$, displacement $\alpha $ and squeezing phase $\theta $ is analyzed, and it can behave a continuous variation from anti-bunching effect to super-bunching effect. The effects of background noise $\gamma $ and detection efficiency $\eta $ on the $g^{(n)}$ are also taken into account. The higher-order coherences $g^{(n)}$ ($n>2$) are more robust against background noise and detection efficiency than the second-order coherence $g^{(2)}$. As the squeezing phase $\theta $ increases from 0 to $\pi $, the $g^{(n)}$ of the phase-dependent squeezed state changes from the minimum $g_{\min }^{(n)}$ ($g_{\min }^{(n)}<1$) to the maximum $g_{\max }^{(n)}$ ($g_{\max }^{(n)}>n!$). It is also verified with experimentally feasible parameters. For $\theta =0$, $\gamma =10^{-5}$, $\eta =0.5$, the strong anti-bunching effects of $g^{(2)}=0.042$, $g^{(3)}=0.010$, $g^{(4)}=0.003$ can be achieved at weak squeezing and small displacements. For $\theta =\pi $ with the same feasible parameters, the super-bunching effects of $g^{(2)}=3.786$, $g^{(3)}=6.190$, $g^{(4)}=375.9$ can be observed at small $r$ and $\alpha $. The results indicate that tuning squeezing phase $\theta $ plays an active role in continuously controlling high-order coherence. This study will contribute to the ongoing attempts to boost the required sensitivity in quantum metrology.

\section*{Appendix}

For an input state, $m$ photons of the total $n$ photons pass through the beam splitter $B0$ with a transmittance of $\eta $, and the photon transmission probability $\Lambda _{m}$ is proportional to the product of $\eta ^{m}$ and $(1-\eta )^{n-m}$, i.e. $\Lambda _{m}\propto $\ $\eta ^{m}(1-\eta )^{(n-m)}$. Meanwhile, it is not known which $m$ photons of the total number $n$ are transmitted through $B0$, so the photon transmission probability must have a binomial combination:
\begin{equation}
\Lambda _{m}=\binom{n}{m}\eta ^{m}(1-\eta )^{(n-m)}.
\end{equation}

Then the photon transmission probability $\Lambda _{m}$ is multiplied by the original photon number distribution of the input state $P_{in}$, and summed over $n$ photon numbers. The photon number distribution after $B0$ can be expressed as a Bernoulli distribution \cite{Scully97}:
\begin{equation}
\Lambda _{m}^{n}=\sum_{n=m}^{\infty }P_{in}(n)\binom{n}{m}\eta ^{m}(1-\eta
)^{(n-m)}.
\end{equation}

Subsequently, the input state is mixed with background noise $\left\vert \beta \right\rangle $ at the beam splitter $B1$. The mixed photon number distribution includes the photon number distribution of the detection efficiency $\Lambda _{m}^{n}$ and the photon number distribution of the background noise $P_{noise}(n)$. Accordingly, the mixed photon number distribution can be expressed as Eq. (3). In this case, $L$ represents the number of photons before $B1$. $N$ photons are transmitted and $L-N$ photons are reflected after $B1$. Then $N$ photons are split by $B2$, and $K$ photons arrive at the detector $D1$, and $N-K$ photons arrive at the detector $D2$. Meanwhile, $L-N$ photons pass through $B3$ and $M$ photons are transmitted into the $D3$, and $L-N-M$ photons are reflected into the $D4$. Since the four detectors are all on-off single-photon counters and $B1$, $B2$, $B3$ are all $50/50$ lossless beam splitters. Therefore, five joint photon probabilities can be obtained through the extended HBT scheme
\begin{eqnarray}
\Gamma _{click\otimes 0} &=&\Gamma (0,0,0,0)=\Lambda _{L=0}^{m},  \nonumber
\\
\Gamma _{click\otimes 1} &=&\Gamma (1,0,0,0)+\Gamma (0,1,0,0)+\Gamma
(0,0,1,0)+\Gamma (0,0,0,1)  \nonumber \\
&=&4\sum_{L=1}^{\infty }\Lambda _{L}^{m}\left( \frac{1}{2}\right) ^{2L}, \nonumber \\
\Gamma _{click\otimes 2} &=&\Gamma (1,1,0,0)+\Gamma (1,0,1,0)+\Gamma
(1,0,0,1)+\Gamma (0,1,1,0) \nonumber \\
&&+\Gamma (0,1,0,1)+\Gamma (0,0,1,1)  \nonumber \\
&=&6\sum_{L=2}^{\infty }\Lambda _{L}^{m}\sum_{k_{1}=1}^{L-1}L!\left( \frac{1%
}{2}\right) ^{2L}/\prod\limits_{i=1}^{2}k_{i}!, \\
\Gamma _{click\otimes 3} &=&\Gamma (1,1,1,0)+\Gamma (1,1,0,1)+\Gamma
(1,0,1,1)+\Gamma (0,1,1,1)  \nonumber \\
&=&4\sum_{L=3}^{\infty }\Lambda
_{L}^{m}\sum_{k_{1}=2}^{L-1}\sum_{k_{3}=1}^{k_{1}-1}L!\left( \frac{1}{2}%
\right) ^{2L}/\prod\limits_{i=2}^{4}k_{i}!, \nonumber \\
\Gamma _{click\otimes 4} &=&\Gamma (1,1,1,1)=\sum_{L=4}^{\infty }\Lambda
_{L}^{m}\sum_{k_{1}=2}^{L-2}\sum_{k_{3}=1}^{k_{1}-1}%
\sum_{k_{5}=1}^{k_{2}-1}L!\left( \frac{1}{2}\right)
^{2L}/\prod\limits_{i=3}^{6}k_{i}!,  \nonumber
\end{eqnarray}
where $k_{1}=N$, $k_{2}=L-N$, $k_{3}=K$, $k_{4}=N-K$, $k_{5}=M$, and $k_{6}=L-N-M$.

Based on the double HBT scheme, the second-order coherence of the input state is determined, and the four detectors can be divided into two groups. $\Gamma (n_{1},n_{2})$ is the joint distribution probability of $n_{1}$ and $n_{2}$ photons detected by the two groups of detectors. Thus, the second-order coherence can be expressed as
\begin{eqnarray}
g^{(2)} &=&\frac{\left\langle n_{1}n_{2}\right\rangle }{\left\langle
n_{1}\right\rangle \left\langle n_{2}\right\rangle }=\frac{\left\langle
n_{1}n_{2}\right\rangle }{\left[ \frac{1}{2}\left\langle n\right\rangle %
\right] ^{2}}=\frac{\sum_{n_{1}n_{2}}n_{1}n_{2}\Gamma (n_{1},n_{2})}{\frac{1%
}{4}[\sum_{n}n\Gamma _{click\otimes n}]^{2}} \nonumber \\
&=&\frac{\Gamma (1,1)+2\Gamma (1,2)+2\Gamma (2,1)+4\Gamma (2,2)}{\frac{1}{4}%
\left\langle n\right\rangle ^{2}} \\
&=&\frac{8\Gamma _{click\otimes 2}+24\Gamma _{click\otimes 3}+48\Gamma
_{click\otimes 4}}{3\left\langle n\right\rangle ^{2}}. \nonumber
\end{eqnarray}

By picking three or four detectors in the double HBT scheme, the third-order or fourth-order coherence of the input state can be obtained as
\begin{eqnarray}
g^{(3)} &=&\frac{\left\langle n_{1}n_{2}n_{3}\right\rangle }{\left\langle
n_{1}\right\rangle \left\langle n_{2}\right\rangle \left\langle
n_{3}\right\rangle }=\frac{\sum_{n_{1}n_{2}n_{3}}n_{1}n_{2}n_{3}\Gamma
(n_{1},n_{2},n_{3})}{\left[ \frac{1}{4}\left\langle n\right\rangle \right]
^{3}}   \nonumber \\
&=&\frac{\Gamma (1,1,1)}{\left[ \frac{1}{4}\left\langle n\right\rangle %
\right] ^{3}}=\frac{\Gamma (1,1,1,0)}{\left[ \frac{1}{4}\left\langle
n\right\rangle \right] ^{3}}=\frac{16\Gamma _{click\otimes 3}}{\left\langle
n\right\rangle ^{3}},
\end{eqnarray}
\begin{eqnarray}
g^{(4)} &=&\frac{\left\langle n_{1}n_{2}n_{3}n_{4}\right\rangle }{%
\left\langle n_{1}\right\rangle \left\langle n_{2}\right\rangle \left\langle
n_{3}\right\rangle \left\langle n_{4}\right\rangle }=\frac{%
\sum_{n_{1}n_{2}n_{3}n_{4}}n_{1}n_{2}n_{3}n_{4}\Gamma
(n_{1},n_{2},n_{3},n_{4})}{\left[ \frac{1}{4}\left\langle n\right\rangle %
\right] ^{4}}  \nonumber \\
&=&\frac{\Gamma (1,1,1,1)}{\left[ \frac{1}{4}\left\langle n\right\rangle %
\right] ^{4}}=\frac{256\Gamma _{click\otimes 4}}{\left\langle n\right\rangle
^{4}}.
\end{eqnarray}

\begin{backmatter}
\bmsection{Funding}
National Natural Science Foundation of China (61875147, 62175176, 62075154, 61731014); Key Research and Development Program of Shanxi Province (International Cooperation, 201903D421049); Shanxi Scholarship Council of China (HGKY2019023); Scientific and Technological Innovation Programs of Higher Education Institutions in Shanxi (201802053, 2019L0131).

\bmsection{Disclosures}
The authors declare that there are no conflicts of interest related to this article.

\bmsection{Data Availability Statement}
Data underlying the results presented in this paper may be obtained from the authors upon reasonable request.

\end{backmatter}


\end{document}